\documentclass[apj]{emulateapj}
\shortauthors{M. Farhang and N. Khosravi}
\usepackage{color}
\usepackage[colorlinks=true,
            linkcolor=black,
            urlcolor=black,
            citecolor=blue, breaklinks=true]{hyperref}
\usepackage{url}
\usepackage{amsmath}
\usepackage{amssymb}
\usepackage{gensymb}
\usepackage{floatrow}

\usepackage{graphics}
\usepackage{graphicx}
\usepackage{natbib}
\input{epsf}
\usepackage{ae,aecompl}
\usepackage{booktabs}

\newcommand{\be}{\begin{equation}}
\newcommand{\ee}{\end{equation}}

\begin{document}
\title{Suggestion: Effective Gravitational Phase Transition:\\ implications for external and internal CMB tensions}
\title{Phenomenological Gravitational Phase Transition in the late Universe and cosmological tensions}
\title{Phenomenological Gravitational Phase Transition:\\ reconciliation between the late and early Universe}

\author{Marzieh Farhang and Nima Khosravi}
\affil{\scriptsize
{Department of Physics, Shahid Beheshti University, 1983969411,  Tehran Iran}}

\email{m_farhang@sbu.ac.ir; n-khosravi@sbu.ac.ir }

\begin{abstract}
In this work we investigate whether a certain phenomenological  extension to the general relativity (GR), in the form of a gravitational phase transition (GPT) in the Universe, can reduce the external {\it Planck} tensions with the local Hubble measurements  and the distribution of matter, characterized by $\sigma_8$, as well as its internal inconsistencies in the lensing amplitude and the low--high $\ell$ parameter estimates. We introduce new degrees of freedom  into the background and the two scalar perturbation equations in the Newtonian gauge, with  simultaneous transitions from an early gravitational phase equivalent to GR toward a late phase. We   model the transition as a \texttt{"tanh"} parametrized by the transition redshift $z_{\rm t}$ and width $\alpha$, with amplitudes $\Lambda(z)$ and $(\mu(z),\gamma(z))$ for the background and perturbations respectively.  
We verify the consistency of the datasets used in this work in the GPT framework and confirm that the individual tensions do not require conflicting transitions.
We find that the joint datasets prefer a recent transition at $z_{\rm t}\approx 0.9$ in the background and perturbed Einstein equations, 
driven mainly by the local Hubble measurement.  This transition relaxes all the tensions considered in this work.  
\end{abstract}

\section{Introduction and Motivations}
The standard model of cosmology, aka $\Lambda$CDM, got well established during the golden era of cosmology, thanks to the high precision data, in particular the fluctuations in the temperature and polarization of the cosmic microwave background radiation (CMB) \citep{pl18}. 
Recently, however, several recent independent observations have put the vanilla $\Lambda$CDM model under tension. 
The most significant is the $H_0$ tension which emerged with the precise local measurements of the Hubble constant, $H_0=74.03\pm 1.42$ km/s/Mpc \citep{R19,R18,R16}. 
This value is in more than $4\sigma$ tension with the $\Lambda$CDM prediction of {\it Planck}, $H_0=67.36 \pm 0.54$ \citep{pl18}. The tension is supported by other independent local measurements \citep{won19,fre19,hua19,khe20,pes20}.
A milder  $2\sigma-3\sigma$ tension is also reported between late measurements of the current matter distribution, as inferred from {\it Planck}, $S_8=\sigma_8\sqrt{\Omega_m/0.3}=0.834\pm0.016$, and the measurements by  DES  $S_8=0.792 \pm 0.024$  \citep{Abbott:2017wau} and by KiDS $S_8=0.766^{+0.020}_{-0.014}$  from weak gravitational lensing and redshift space distortion \citep{hey20}.

Besides these tensions, there are also reports on internal discrepancies in the {\it Planck} temperature dataset when interpreted in the $\Lambda$CDM framework \citep{pl18}.
 The first is the $\sim 2 \sigma$ deviation of the lensing amplitude, $A_{\rm lens}$, from unity which is required by the theoretical consistency of the theory, and is also consistent with the $A_{\rm lens}$ estimation from the {\it Planck} four point correlation function.  
 The second is the inconsistency of the measurements of certain cosmological parameters, derived independently from the low and high CMB multipoles\footnote{This inconsistency may be related to the former anomaly in the lensing amplitude \citep{pl18}.}.
To make this list of CMB-based cosmic curiosities complete, we also mention that there are also the broadly addressed CMB spatial anomalies which challenge the isotropy and homogeneity of the Universe such as the dipole modulation, the quadrupole-octopole alignment and the CMB cold spot \citep{pl19stat} . 

There have been extensive studies of all these temporal and spatial tensions and anomalies, in particular, of the possibility of them being produced by yet  unexplained systematics.
Moreover, many recent  beyond--$\Lambda$CDM proposals have emerged by relaxing some of its  assumptions.
These include treating the neutrino masses as free parameters, however, found to increase the $\sigma_8$ tension \citep{pl18}, tweaking the physics of the early Universe \citep{kee20}, and to look for windows in the dark side of the Universe, in both  dark matter \citep[see, e.g.,][]{vat19})  and dark energy sectors, \citep[see, e.g.,][]{kee19,dival20,Zhao:2017cud,Khosravi:2017hfi,Yan:2019gbw,Braglia:2020iik,Raveri:2019mxg,Poulin:2018cxd,DiValentino:2017iww,Gomez-Valent:2020mqn,luc20,yan18,Yang:2018uae,DiValentino:2019ffd,DiValentino:2019jae,Li:2019yem}.
The physics of critical phenomena is also a recent proposal to support the physical motivation for a phase transition in the behavior of dark energy  and is phenomenologically  explored at the background level \citep{Banihashemi:2018has,Banihashemi:2018oxo}.

In this paper we investigate whether 
a gravitational phase transition, referred to GPT, can  address the above internal and external tensions. 
For this purpose, we explore the gravitational parameter space from an effective viewpoint. Einstein's intuition in general relativity (GR) was absolutely genius.  Yet, a more systematic approach to the theory of  gravitation is  the formalism of the effective field theory (EFT) 
which includes all theoretically possible terms, e.g., all those satisfying a required symmetry \citep{Gubitosi:2012hu,Bloomfield:2012ff}. 
In this framework GR can be considered as a special case, obtained through the  fine-tuning of some of the EFT parameters. 
GR is also the most economic formulation in terms of the minimum number of degrees of freedom (dof's) required for a gravitational theory. Any modification to it will therefore propagate new dof's. 
In the EFT approach, the modification to the background gravity can be modeled through replacing the cosmological constant by a time-dependent field\footnote{The Newton constant $G_{\rm N}$ is also a possibility. For our purpose, however, it is degenerate with the other parameters.}, 
while two new scalar dof's are required to describe modification to perturbation equations in the linear level in the Newtonian gauge.
 
In this work we parametrize these dof's by the three functions\footnote{Note that in principle these variables can be scale dependent as $\mu(k,z)$ and $\gamma(k,z)$.} $\Lambda(z)$ and $\left(\mu(z),\Sigma(z)\right)$, assumed to be constant in $\Lambda$CDM with   $\Lambda(z)=\Lambda$ and $(\mu(z),\Sigma(z))=(1,-1)$.
Here we parametrize their redshift dependence by a \texttt{"tanh"}. For this assumption, we are motivated by the idea of a possible phase transition in the dark energy, proposed to relax the cosmological tensions\footnote{In this work we do not address the spatial CMB anomalies. However, we  showed in \cite{Banihashemi:2018has}  that a phase transition inspired by the physics of  critical phenomena can provide a rich framework
for simultaneous exploration of spatial anomalies and temporal tensions.}. We use \texttt{"tanh"} to phenomenologically model this phase transition in the mean field approximation. 
We explore the space spanned by the new GPT dof's to investigate the consistency of the various cosmological datasets in this framework and to investigate the power of this proposal in relaxing the tensions. 

In the rest of this paper we first introduce the phenomenological effective framework of GPT (Section~\ref{sec:model})  and  the datasets (Section~\ref{sec:dat}). The results on the resolution of the external and internal tensions are presented in Sections~\ref{sec:exttensions} and \ref{sec:inttensions} and we conclude in Section~\ref{sec:conclusion}. 
%------------------------------------------------------------
\section{The GPT model}\label{sec:model}
%------------------------------------------------------------
In this work we propose a single late--time transition in the behavior of the gravitational theory, modeled phenomenologically by a \texttt{"tanh"} function. By late time we refer to any epoch between the last scattering surface and now. 
We assume the early-time gravity, well before the transition, is described by GR. This assumption could be relaxed in principle in the cost of increasing the number of free parameters of the theory.

We assume our Universe is isotropic, homogeneous and  flat, with the background evolution described by  
\begin{eqnarray}
H^2(z)=H_0^2 \bigg[\Omega_r (1+z)^4+\Omega_m (1+z)^3+\Omega_\Lambda(z)\bigg]
\end{eqnarray}
and with linear scalar perturbations described by 
\begin{eqnarray}
ds^2=a^2(\tau)\left[-(1+2\Psi)d\tau^2+(1-2\Phi)dr^2\right]
\end{eqnarray}
in the Newtonian gauge. Here $a(\tau)$ is the scale factor and $\Psi$ and $\Phi$ characterize the scalar perturbations.
Phenomenologically, the modified Einstein equations in Fourier space describing the evolution of perturbations in the linear level are given by 
%-----------------------------------------------------------
\begin{eqnarray}\label{eq:mu}
&&k^2 \Psi = -\mu(z)\,4\pi G a^2\,\left[\rho\Delta+3(\rho+P)\sigma\right]\\
&&k^2\left[\Phi-\gamma(z)\Psi\right]=\mu(z)\,12\pi G a^2\,(\rho+P)\sigma, \label{eq:gamma}
\end{eqnarray}
%-------------------------------------------------------------
where $\mu(z)$ and $\gamma(z)$ encode all the information about the departure of the gravitational theory from Einstein gravity.
Alternatively,  Equation~(\ref{eq:gamma}) can be replaced by  
 \begin{equation}\label{eq:lensing}
 k^2(\Phi+\Psi)=8\pi G a^2 \,\Sigma(z)\, \rho\, \Delta
\end{equation}
which is particularly useful for characterizing the gravitational lensing. 

We model the modification to the background by a transition in the cosmological constant 
%-----------------------------------
\begin{eqnarray}\label{eq:delv}
\Omega_\Lambda(z)=\Omega^{\rm early}_\Lambda+\Delta_\Lambda\,\frac{1+\tanh\big[\alpha\,(z_{\rm t}-z)\big]}{2}.
\end{eqnarray}
%------------------------------------
where $z_{\rm t}$ is the transition redshift and $\alpha$ is the inverse of transition's width. 
Flatness requires $\Omega_{\rm r}+\Omega_{\rm m}+\Omega_\Lambda=1$, where 
$\Omega_\Lambda\equiv\Omega_\Lambda(z=0)=\Omega^{\rm early}+\Delta_\Lambda(1+\tanh(\alpha z_{\rm t}))/2$. 
Similarly, the modifications to the perturbation equations are characterized by transitions in $\mu(z)$ and $\gamma(z)$,
%---------
\begin{eqnarray}\label{eq:pert}
\mu(z)&=&\mu^{\rm early}+\Delta_{\mu} \frac{1+\tanh\big[\alpha\,(z_{\rm t}-z)\big]}{2}, \nonumber \\
\gamma(z)&=&\gamma^{\rm early} + \Delta_\gamma \frac{1+\tanh\big[\alpha\,(z_{\rm t}-z)\big]}{2}.
\end{eqnarray}
%----------------------------
 where $\mu^{\rm early}=\gamma^{\rm early}=1$ and $\Sigma^{\rm early}=-1$.
%----------------------------------------------
\section{Analysis and datasets}\label{sec:dat}
%------------------------------------------------------------------------------------------
%
In this work we use the  CMB temperature and $E$-mode polarization power spectra as measured by {\it Planck} 2018 data, the same as  the {\it reference likelihood} in the \cite{pl18}. We refer to it as Pl18. In the final analysis of this paper in section~\ref{sec:conclusion}, where the results for the joint datasets are presented,  we also include the {\it Planck} CMB lensing, labeled as + lensing.

As the primary late time data point we use the most recent Hubble constant measurement from \cite{R19}, given by $H_0=74.03 \pm 1.42$ km/s/Mpc, referred to as R19 throughout the paper. 
We also use the cosmic shear and galaxy clustering measurements, as well as their cross correlation,  from the first year observation of the Dark Energy Survey \citep{Abbott:2017wau},  referred to as DES in the following sections. 
In some cases we assume a futuristic observational scenario  for $H_0$ measurement, referred to as H073,  where the current  best-fit value of $H_0$ is retained with much tighter bounds.  

The parameter space in this work consists of the standard cosmological parameters, $(\Omega_{\rm b}h^2,\Omega_{\rm c}h^2, \theta, A_{\rm s}, n_{\rm s}, \tau)$, and the parameters effectively describing the  GPT, $(z_{\rm t},\alpha,\Delta_\Lambda,\Delta_\mu, \Delta_\gamma)$ (see Equations~\ref{eq:delv} and \ref{eq:pert}). 
 In section~\ref{sec:lens}, $A_{\rm lens}$ is also set free as part of the study of the internal consistency of  the {\it {\it Planck}} temperature dataset in the GPT framework.
We use the publicly available Markov-Chain Monte-Carlo (MCMC) code, CosmoMC\footnote{https://cosmologist.info/cosmomc/}, to sample the parameter space.
%-----------------------------------------------------------------------------------------
\section{results I: $H_0$ and $\sigma_8$ tensions} \label{sec:exttensions}
%------------------------------------------------------------------
In this section we explore the potential of the GPT  parameter space  for peaceful explanation of the tensions between CMB-based inferences and R19 measurements of $H_0$ (Section~\ref{sec:h0}) and DES implications for the growth of structures (Section~\ref{sec:s8}). 
%----------------------------------
\subsection{Hubble tension}\label{sec:h0}
\begin{figure}[t]
    \centering
    \includegraphics[scale=0.7]{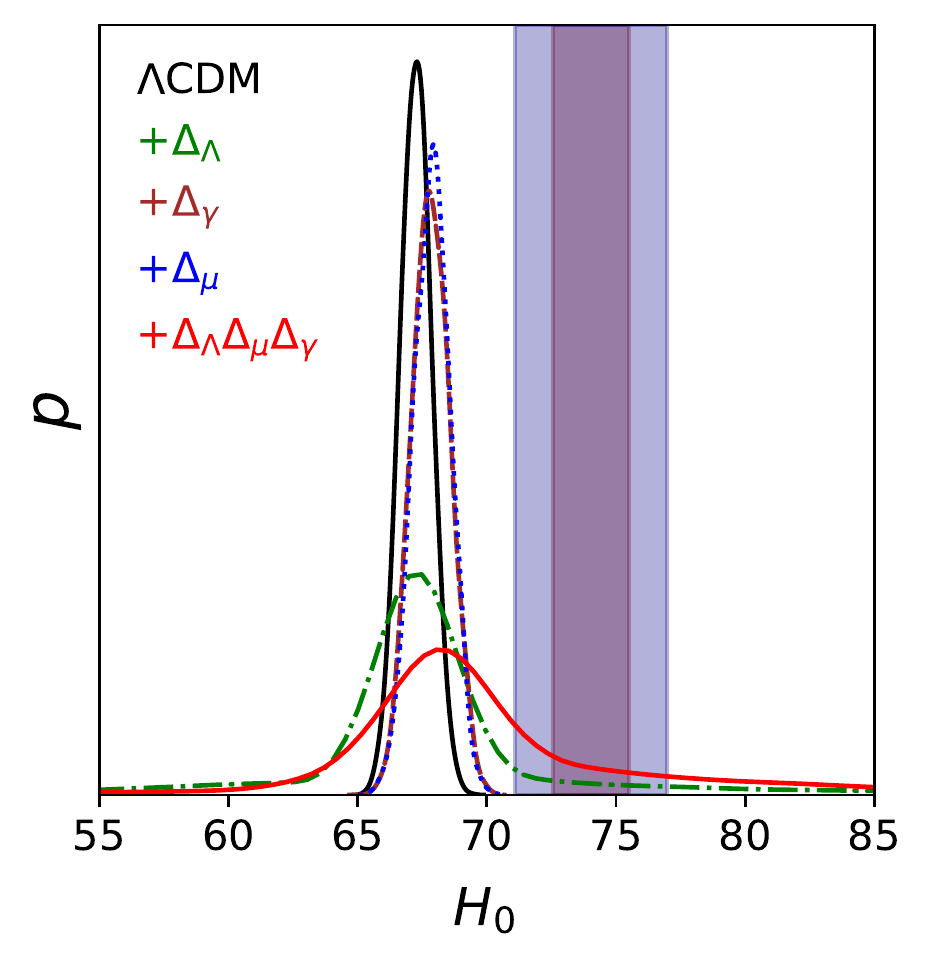}    
    \caption{The $1$D marginalized $H_0$ likelihood as measured by Pl18 compared for the $\Lambda$CDM model  and the GPT minimal ($+\Delta_\Lambda$, $+\Delta_\mu$ and $+\Delta_\gamma$) and full ($+\Delta_\Lambda\Delta_\mu\Delta_\gamma$) scenarios. The tension with R19 measurements (shaded purple $1-2\sigma$ region) is relaxed with transitions in the background and perturbation dof's ($+\Delta_\Lambda\Delta_\mu\Delta_\gamma$). }
    \label{fig:h0}
\end{figure}

%----------------------------------
Figure~\ref{fig:h0} compares the Pl18 $1$D marginalized likelihoods for $H_0$ in the the $\Lambda$CDM and GPT scenarios.
The minimal GPT scenarios, with a single GPT amplitude allowed to vary, correspond to background-only ($+\Delta_\Lambda$) and perturbation-only transitions ($+\Delta_{\gamma}$ and $+\Delta_{\mu}$). 
The full model, labeled as $+\Delta_{\Lambda}\Delta_{\gamma}\Delta_{\mu}$, includes simultaneous transitions in the background and the two perturbed Einstein equations. 
The shaded vertical bars indicate the $1\sigma$ and $2\sigma$ confidence regions of $H_0$ by R19.
We see that
the background  transition opens up the Pl18 constraint on $H_0$ and reduces the tension. 
Transition in a single perturbation equation, on the other hand,  by
either $\Delta_{\mu}$ or $\Delta_{\gamma}$, slightly shifts the position of the $H_0$ likelihood peak without significant 
impact on the distribution width. 
Allowing for simultaneous transitions in the background and perturbations leads to considerable increase in the the bounds on $H_0$ and removes the tension. 

%-------------------------------------------------------------------------------------------
%------------------------------------------------------------------------------------------

This relaxation  in the full GPT parameter space is not unexpected and can be explained as follows.  
The tight observational constraint on the angular size of the sound horizon, 
  $\theta_{\rm s}=r_{\rm s}/D_{\rm LSS}$, where $r_{\rm s}= \int_{z_{\rm drag}}^{\infty}c_{\rm s}dz\slash H(z)$ and $D_{\rm LSS}=\int^{z_{\rm LSS}}_{0}dz\slash H(z)$, leads in turn to tight constraints on $\Omega_m h^3$  in  $\Lambda$CDM \citep{Percival:2002gq,pl18}. 
In GPT, with a different expansion history, the degeneracy between the parameters introduced by the new dof's in 
 $\Omega_\Lambda$, would lead to enhanced bounds on $\Omega_m$ and $H_0$, after marginalization over other parameters.
 On the other hand, there are tight constraints on matter density, 
 $\Omega_m h^2$, due to its relatively distinct imprints on temperature power spectrum, through affecting the epoch of equality, the depth of potential wells (and thus the monopole at the last scattering) and the lensing of CMB photons. 
Therefore, the background-only transition by itself is inadequate to explain the observed $H_0=73.4$ km/s/Mpc, and would benefit from the dof's in perturbation equations which are partially degenerate with matter density. 
This in turn would induce correlations with other standard parameters through the early-time imprints of matter density, which cannot be compensated by the late-time imprints of GPT dof's.  

%-----------------------------------------------------------------------------------------
\begin{figure}[t]
\centering
	\includegraphics[scale=0.7]{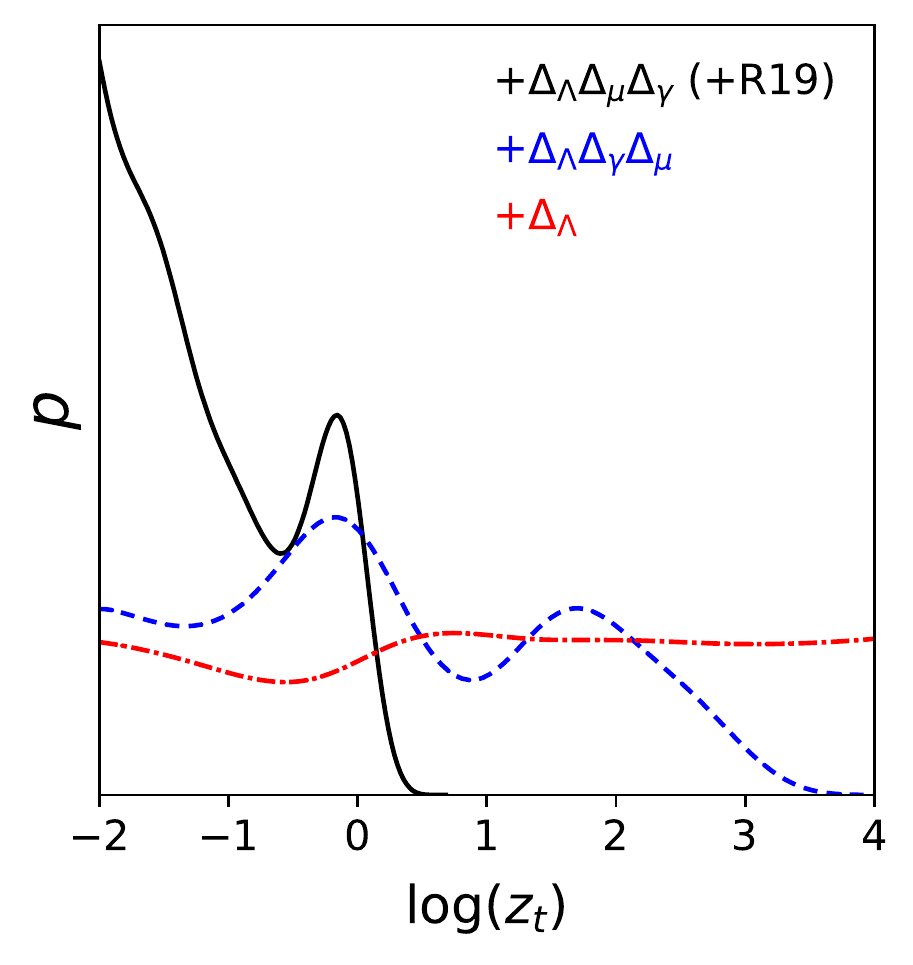}
	\caption{The marginalized likelihood of $\log z_{\rm t}$ as measured by Pl18 for the background and full GPT parameters. The black curve represents the likelihood by the joint Pl18+R19 for the full GPT, and prefers a transition at $z_{\rm t}\approx 0.9$. The $z_{\rm t} \approx 0$ transition is not physically motivated and therefore not further considered here. }\label{fig:logzt}
\end{figure}
%-----------------------------------------------------------------------------------------

Figure~\ref{fig:logzt} illustrates the distribution of the transition redshift as measured by the Pl18 and Pl18+R19  for the background and full GPT scenarios.
The combined dataset prefers a very recent transition at $z\approx0$
and also allows for a transition at $z_{\rm t} \approx 0.9$.
The $z\approx0$ scenario, imposed by the $H_0$ prior from R19,  is the trivial reconciliation of the late $H_0$ with the CMB inference of $H_0$. 
Its corresponding background history respects $\Lambda$CDM until very recently, and make a sudden switch to yield the high $H_0$ of R19. This fine-tuned scenario is however the least physically-motivated. 
We therefore regard the  $z_{\rm t} \approx 0.9$ as the main scenario suggested by data.

The sharp decline in the high-$z$ tail of this transition can be explained as follows. The background history is equivalent to $\Lambda$CDM at high $z$, and the accompanying transitions in  $\mu$ and $\gamma$, with effectively no modification to the background, cannot solve the tension (Figure~\ref{fig:h0}). 
This also explains the non-vanishing probability of  $z_{\rm t}$ at higher redshifts for the CMB-only measurement in the background GPT scenario. 
%-----------------------------------------------------------------------------------------
\subsection{$\sigma_8$ tension}\label{sec:s8}
%-----------------------------------------------------------------------------------------
The claims of moderate tension between the Pl18 $\Lambda$CDM-based inferences and DES probes are mostly apparent in combinations of background matter  density and its fluctuations  \citep[see, e.g., Fig. 20 of][]{pl18}.
We include the GPT parameters in the analysis and find the tension is relieved by the boosted  parameter ranges allowed by the data, 
\begin{eqnarray}
S_8= 0.836\pm0.042 , ~~~~~\Omega_{\rm m}= 0.290\pm0.048 
\end{eqnarray} 
from Pl18, and
\begin{eqnarray}
S_8= 0.696 \pm 0.134, ~~~~~ \Omega_{\rm m}= 0.281\pm 0.053 
\end{eqnarray} 
from DES. 
We also verified the consistency of the GPT parameters as  measured by the individual Pl18 and DES datasets.
 We therefore conclude the consistency of the GPT cosmological inferences from these two datasets.
%-------------------------------------------------------------------------------------------
\section{results II: internal (in)consistencies in {\it {\it Planck}}} \label{sec:inttensions}
%-------------------------------------------------------------------------------------------
Here we explore whether the GPT scenario can explain the reported internal curious features of Pl18 when interpreted in the $\Lambda$CDM framework. These include the parameter shifts when separately inferred from high- and low-$\ell$ regimes and the preference of the temperature power spectrum for stronger lensing than is required by $\Lambda$CDM.  
It is suggested that non-zero spatial curvature can resolve the latter, however, at the cost of increased $H_0$ and $\sigma_8$  tensions \citep{DiValentino:2019qzk}.
 
 %------------------------------------------------------------------------------------------
\begin{figure}
    \centering
    \includegraphics[scale=0.45]{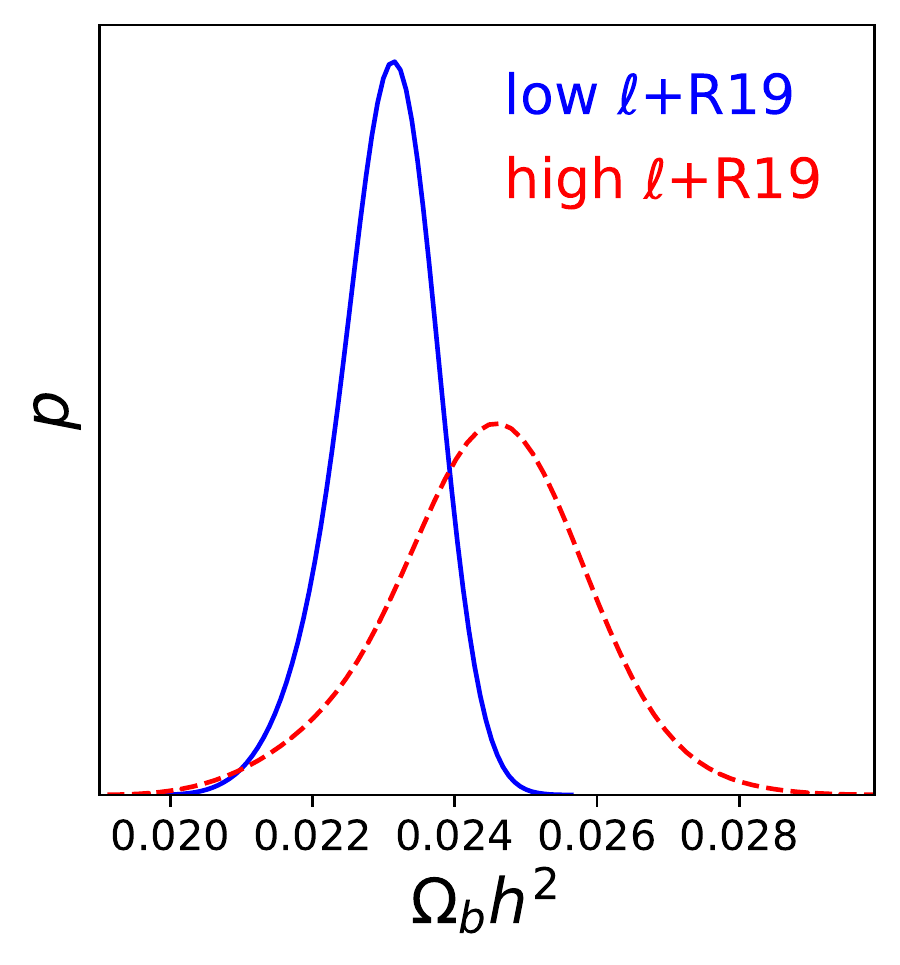}
    \includegraphics[scale=0.45]{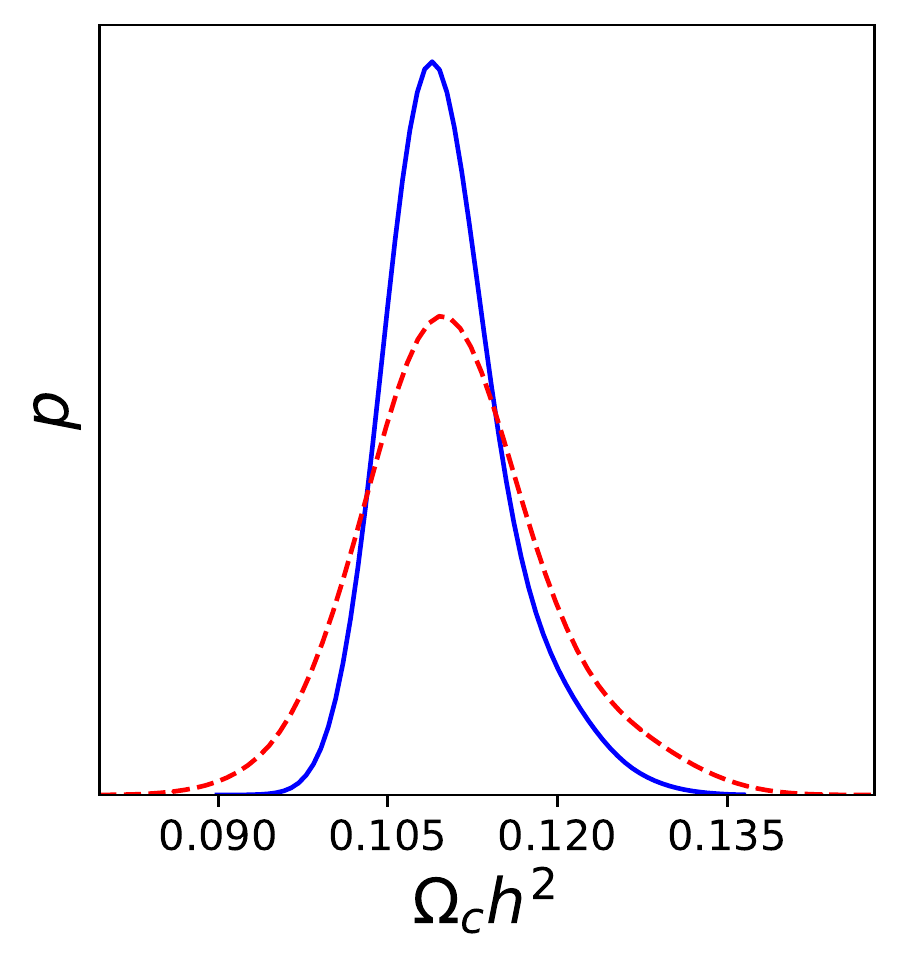} 
       \includegraphics[scale=0.45]{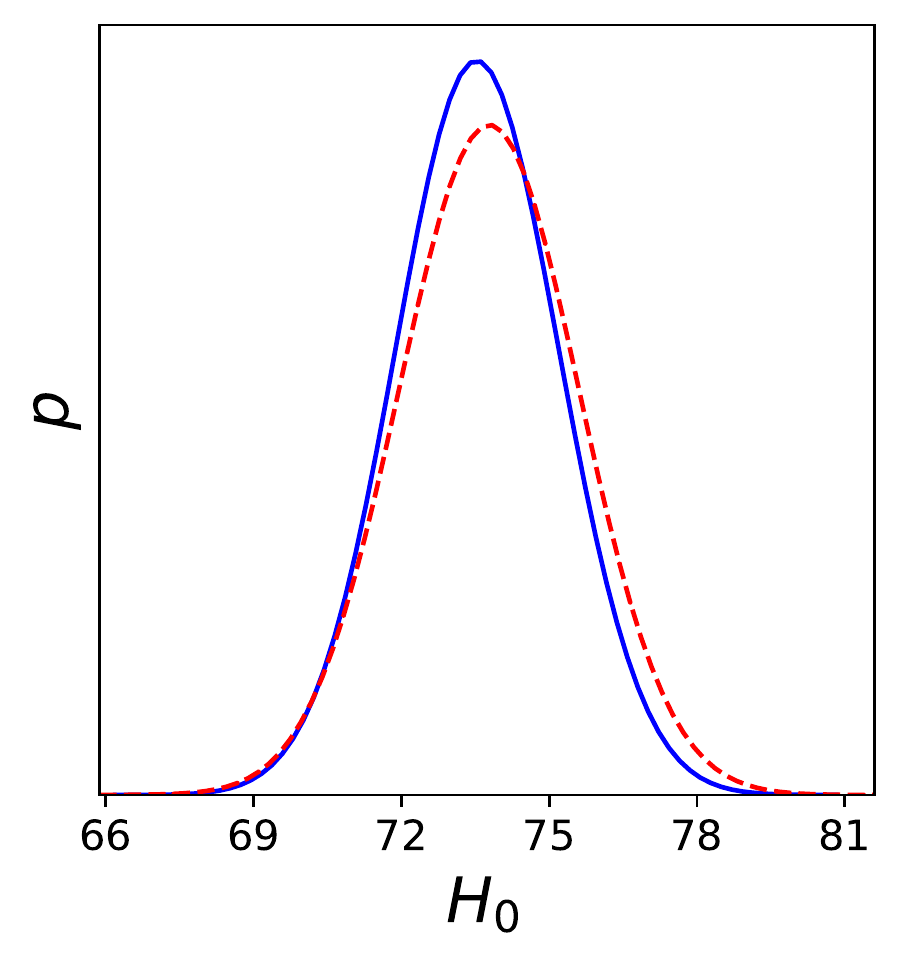}
    \includegraphics[scale=0.45]{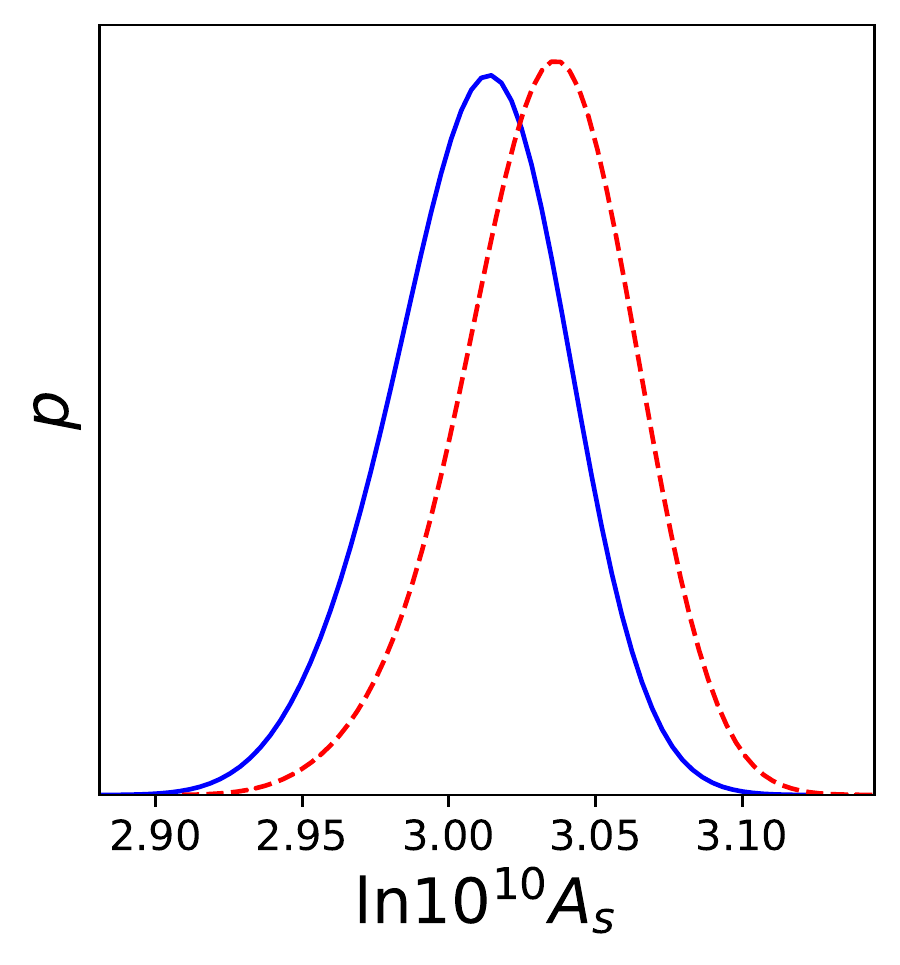}
    \includegraphics[scale=0.45]{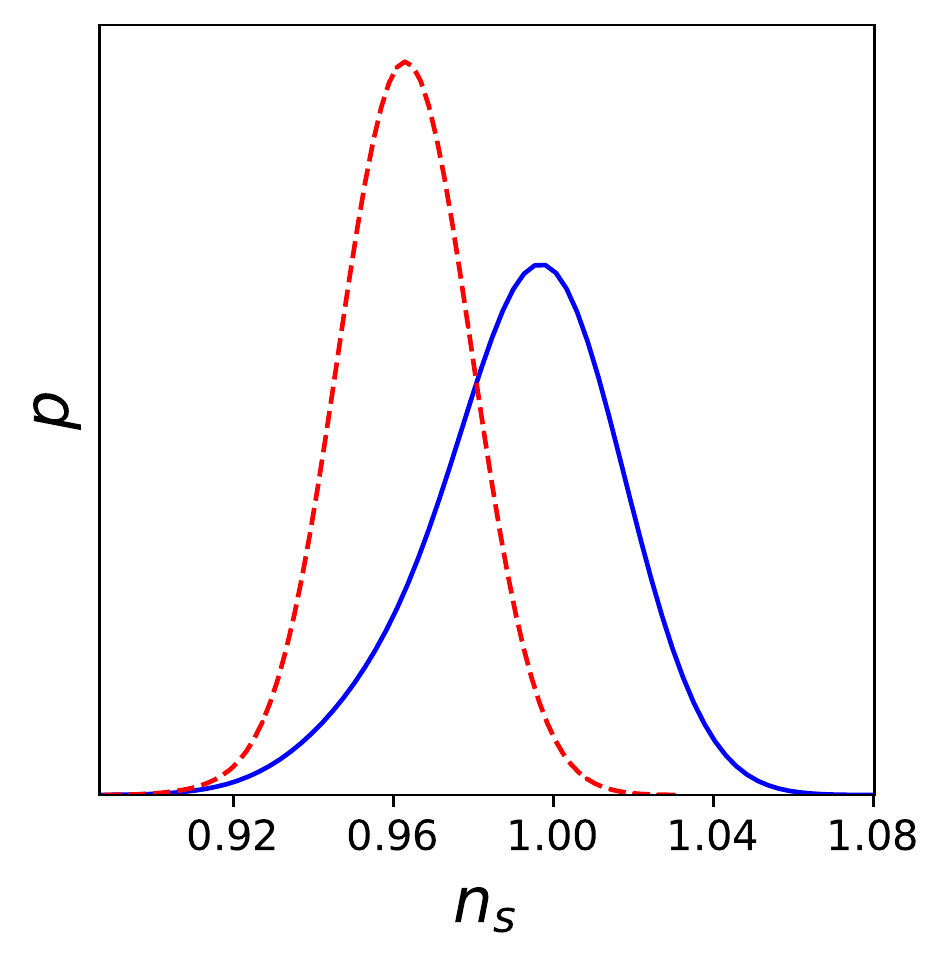}
       \includegraphics[scale=0.45]{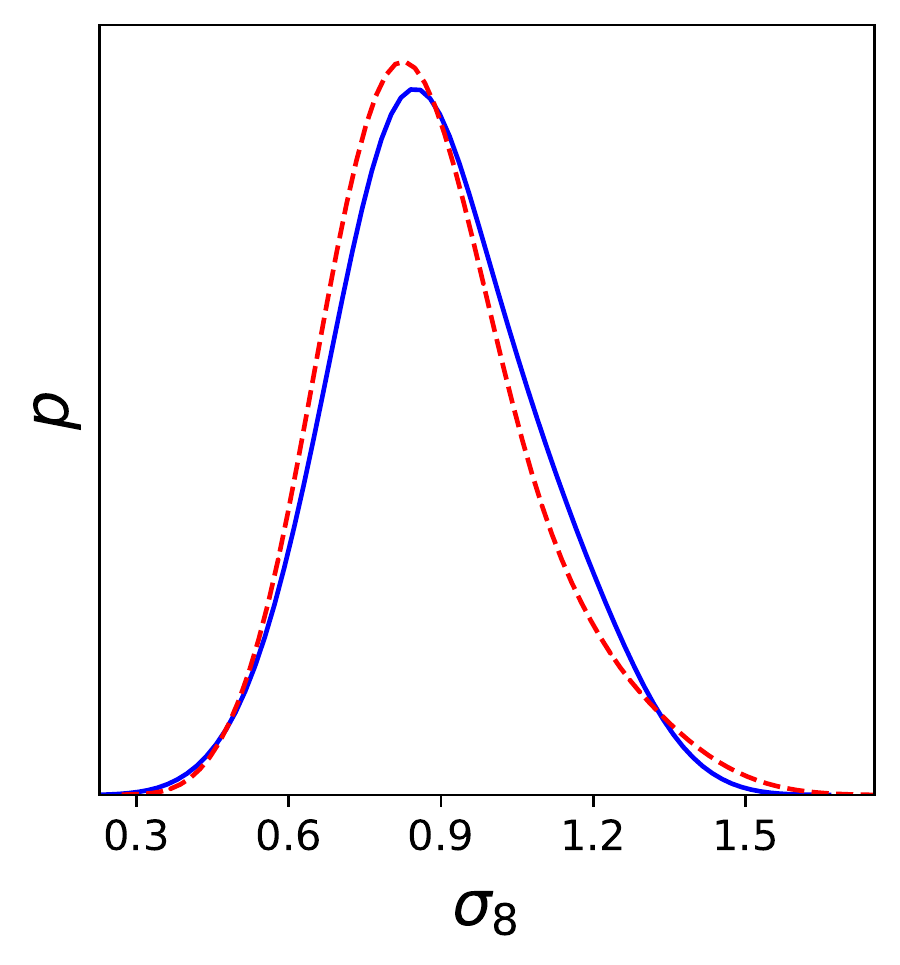}
    \caption{The marginalized likelihoods for various standard cosmological parameters  in the GPT model as measured by low and high $\ell$ {\it Planck} temperature and low-$E$ anisotropies, with the R19 prior on $H_0$. The measurements from the two multipole regimes are found to be consistent.}
    \label{fig:lohi}
\end{figure}

  %-------------------------------------------------------------------------------------------
\subsection{low--high $\ell$ tension}
%-------------------------------------------------------------------------------------------
We  measure the different cosmic parameters in the GPT model separately with the low ($\ell \le 800$) and high ($800< \ell \le 2500$) multipoles of the {\it Planck}  temperature anisotropies. Figure~\ref{fig:lohi} compares the marginalized likelihoods for some of the standard parameters.
In both cases, we also include the R19 (justified by its consistency with Pl18,  see Section~\ref{sec:h0}) and the {\it Planck} low-$E$, as otherwise the GPT parameters and the amplitude of primordial scalar perturbations would be unconstrained by the separate low and high $\ell$ data.
The scalar index is the only parameter with a noticeable shift in its best-fit  values. 
The measurements, however, are still consistent due to the notable overlap of the distribution ranges. 
%------------------------------------------------------------ 
%-------------------------------------------------------------------------------------------

\begin{figure}
    \centering
   \includegraphics[width=4.2cm]{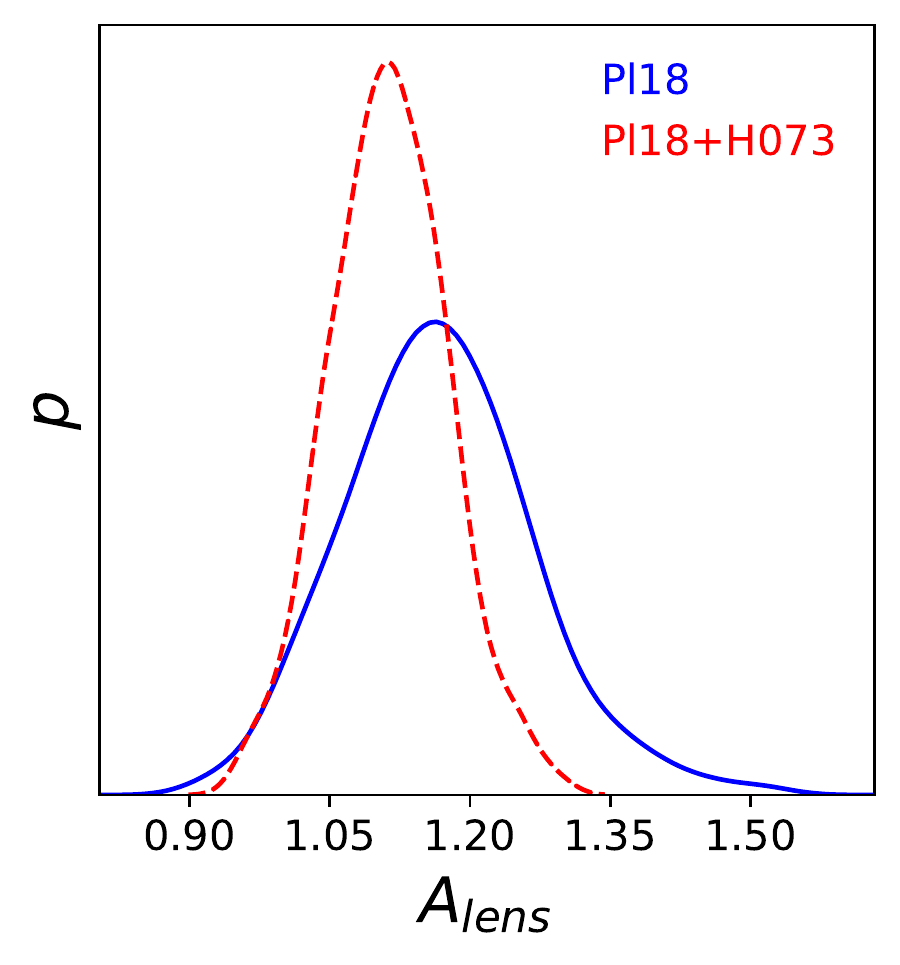}
   \includegraphics[width=4.2cm]{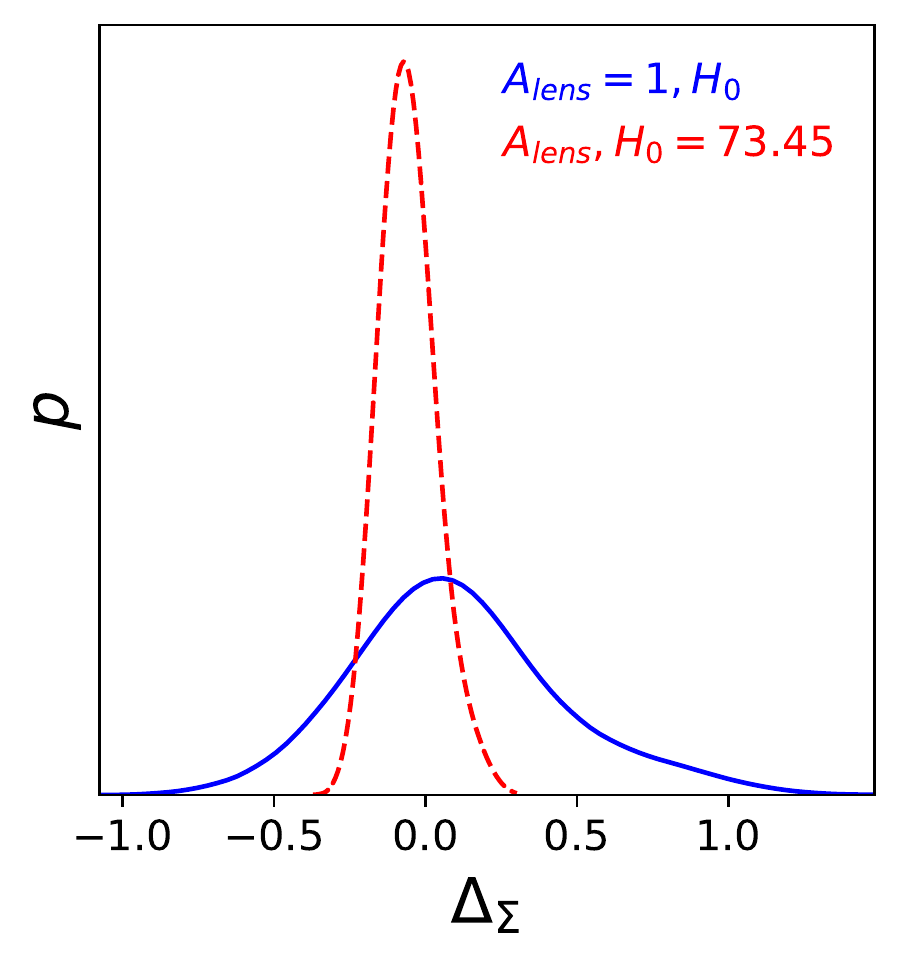}
   \caption{Left: Marginalized likelihoods of the lensing amplitude in the GPT framework, as measured by Pl18 (solid blue) and Pl18+H073 (dashed red). The  $\Lambda$CDM tension of $A_{\rm lens}$ is reduced to less than a $\sigma$ in GPT. The Pl18+H073  curve shows that possible tight future $H_0$ constraints, even if far from the current $\Lambda$CDM inferences, will not reduce the $A_{\rm lens}=1$ consistency in GPT. \\
   Right: Marginalized likelihood for the derived lensing parameter $\Delta_\Sigma$ (see Equation~\ref{eq:Sigmaidentitiy}) compared for two cases, one with $A_{\rm lens} $ free and with a tight prior on $H_0$ through including H073 (red dashed curve)  and the other with $A_{\rm lens}$ set to unity and with the R19 prior on $H_0$ (solid blue curve). We see that the two likelihoods overlap, implying that the required $\Delta_\Sigma$ to get H073 does not conflict with the $\Delta_\Sigma$ needed to ensure $A_{\rm lens}=1$.}
   \label{fig:Al}
\end{figure}

%-------------------------------------------------------------------------------------------
\subsection{lensing tension}\label{sec:lens}
%-------------------------------------------------------------------------------------------
The lensing-specific dof $\Sigma$ in Eq.~\ref{eq:lensing} is related to $\mu$ and $\gamma$ through
\begin{eqnarray}
\Sigma&=&-\frac{\mu}{2}(1+\gamma)\\ \nonumber \label{eq:Sigmaidentitiy}
&\approx&\Sigma^{\rm early} + \Delta_\Sigma \frac{1+\tanh\big[\alpha\,(z_t-z)\big]}{2}
\end{eqnarray}
where $\Delta_\Sigma=-\Delta_\mu-{\Delta_\gamma/}{2}$ and the second line only includes linear contributions  of $\Delta_\mu$ and $\Delta_\gamma$. The goal is to investigate whether modifications to $\Sigma$ in the GPT framework can resolve or reduce the tension in $A_{\rm lens}$. 
The left panel in Figure~\ref{fig:Al}  illustrates the distribution of $A_{\rm lens}$  in GPT  as measured by Pl18, without and with a tight $H_0$ prior, labeled as Pl18 and Pl18+H073 respectively. 
We find that $A_{\rm lens}$ is consistent with one in both cases. The futuristic prior imposed by H073 would tighten the bound and push the measurement toward lower amplitudes, without reducing the $A_{\rm lens}=1$ consistency. We conclude that  the GPT model would continue to explain the Pl18 data if the current high local Hubble value is measured with much higher accuracy.

%-------------------------------------------------------------------------------------------
\begin{figure*}
    \centering
    \includegraphics[scale=0.5]{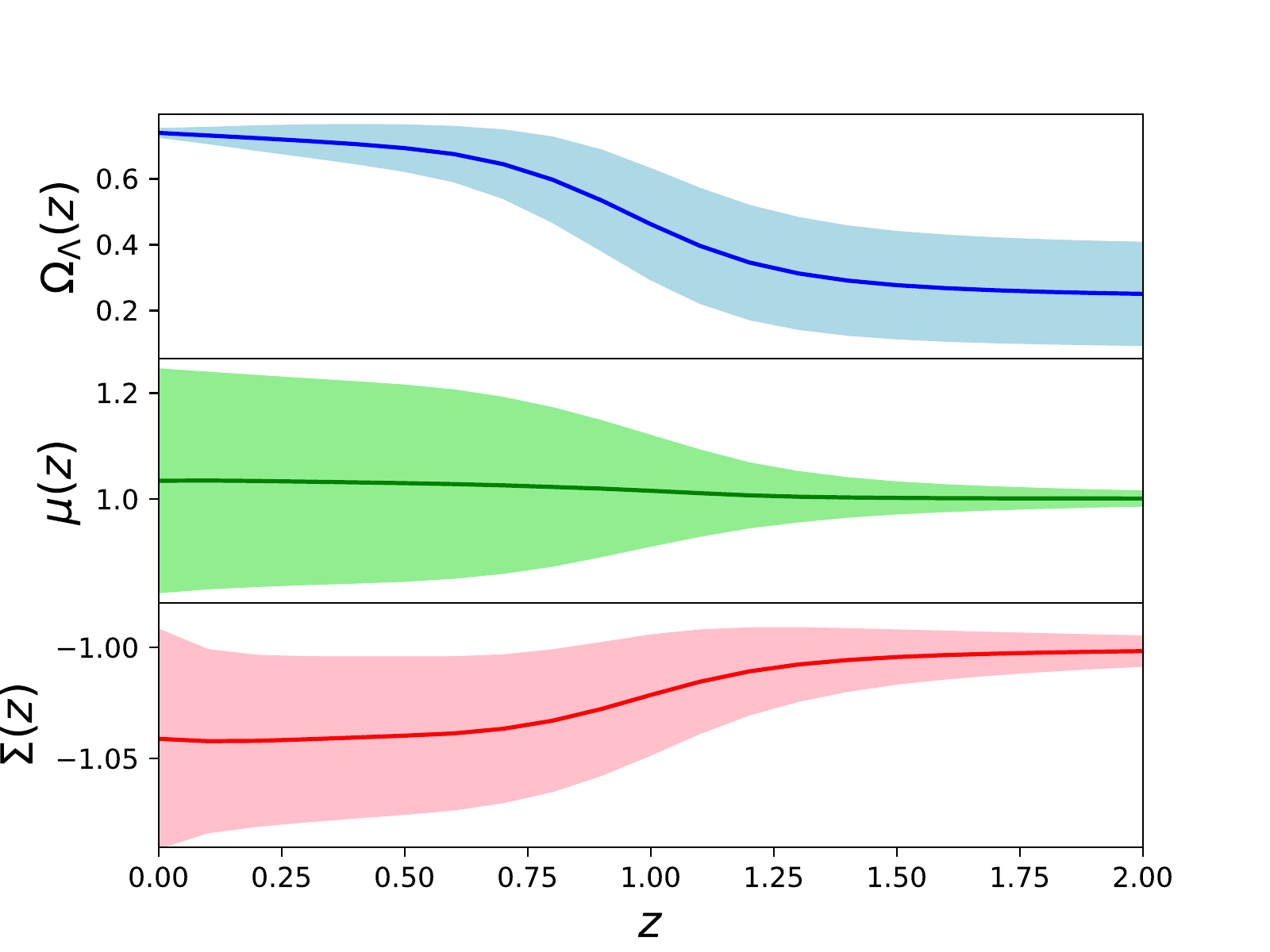}    
    \includegraphics[scale=0.5]{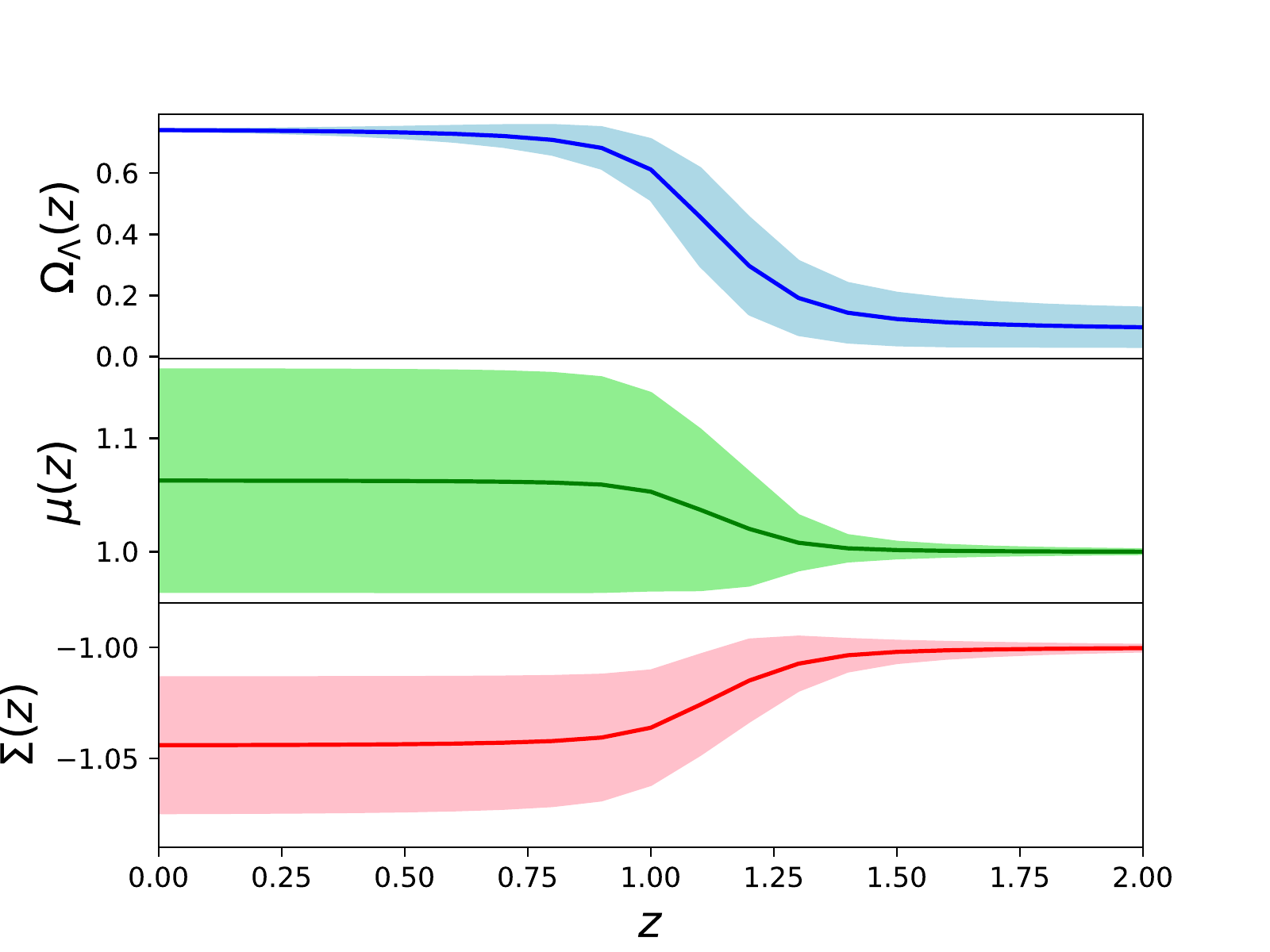}    
    \caption{The $1\sigma$ trajectories in redshift space traced by the various  GPT parameters with the joint datasets Pl18+R19 (left) and  Pl18+lensing+H073+DES (right). These transitions can relax the {\it Planck} external tensions and internal inconsistencies addressed in this work.  The  deviations in $\Omega_\Lambda(z)$ and $\Sigma(z)$ from their $\Lambda$CDM values are mainly imposed by  $H_0$ priors, most vivid in the right panel.} 
    \label{fig:traj}
\end{figure*}
%-----------------------------------------------------------------------------------------
It is also important to verify the consistency of the required lensing perturbations for the independent resolution of the lensing and  Hubble tensions. 
We consider two observational scenarios, each with a tight prior on one of the two parameters $H_0$ and $A_{\rm lens}$, and with the goal to measure the other parameter.  
The right panel of Figure~\ref{fig:Al} compares the likelihoods for the two cases, one with Pl18+H073 with $A_{\rm lens}$ free (the dashed red curve) and one with Pl18 with $A_{\rm lens}=1$ and $H_0$ free (the blue solid curve). The two likelihoods overlap, verifying that the two requirements in their extreme cases can be simultaneously satisfied with agreeing lensing corrections.  
%------------------------------------------- 
 \section{conclusion}\label{sec:conclusion}
 %------------------------------------------- 
  The analysis of the previous sections illustrates that the discussed {\it Planck} external ($H_0$ and $\sigma_8$) tensions and internal (low-high $\ell$ and lensing amplitude) inconsistencies are relaxed in the GPT framework.
 The best-fit trajectories and their $1\sigma$ confidence regions in the left panel of Figure~\ref{fig:traj} show the transitions in $\Omega_\Lambda(z)$, $\mu(z)$ and $\gamma(z)$  required to relax the Hubble tension.
  We have set $A_{\rm lens}=1$ as its consistency with R19 (and H073) was shown in Section~\ref{sec:lens}.
  The best-fit trajectories of all three amplitudes show preference for time variations and smoothly go to their  early-time values.   
  However, the shaded $1\sigma$ regions surrounding the best-fit trajectories show that $\mu$ is fully consistent with GR 
 while $\Sigma$ and $\Omega_\Lambda$ noticeably deviate from their $\Lambda$CDM values. % 
 
It is interesting to see how tighter futuristic
Hubble measurements would constrain the trajectories.
This is shown in the right panel of figure~\ref{fig:traj}, where we have used the full data combination of  Pl18+lensing+H073+DES. 
The lensing consistency and Hubble constraint assumptions mainly impact the 
$\Lambda(z)$ and $\Sigma(z)$ trajectories. 
One expects from Equation~\ref{eq:gamma} that $\mu(z)$ would be best constrained by measurements of matter distribution, characterized, e.g., by tight bounds on  $\sigma_8$. However, these measurements are model-dependent and it is not feasible to directly assume a tight prior on $\sigma_8$ to investigate whether and how it would be compensated by a proper transition in $\mu(z)$. One could, instead, simulate observations (of, e.g., power spectrum of cosmic shear or galaxy clustering) for an assumed set of GPT parameters, and see how those would lead to inconsistencies in the parameter measurements if analyzed in the $\Lambda$CDM framework. We did not explore this here. 

We therefore find a late-time gravitational transition, simultaneous in the dof's of the background  equation and the  perturbed Einstein equations, relaxes the main current external {\it Planck} tensions, as well as its internal inconsistencies. 
The framework of phase transition also opens the room to address the spatial anomaly as was investigated in \cite{Banihashemi:2018has}.
We thus view the GPT scenario as a rich mechanism to  
encompass various datasets and phenomenologically address their temporal and spatial tensions and anomalies. 
In this work our focus was on the three datasets of Pl18(+lensing), R19 and DES. In future analysis one should also include a broader set of data, such as BAO and high-$z$ supernovae.
The next natural extension is to allow for the early phase of gravity to deviate from GR. A main concern would then be the many dof's which, if not dealt with properly, will inflate the parameter space and leave the parameters practically unmeasurable. This will be explored in future work.

\section*{Acknowledgement}
The numerical calculations of this work were carried out on the computing
cluster of the Canadian Institute for Theoretical Astrophysics (CITA), University of Toronto. 

\bibliography{gpt}
\bibliographystyle{apj}

\end{document}